\documentclass{emulateapj}
\usepackage{epstopdf}

\newcommand{\teff}{T$_{eff}$}
\newcommand{\logg}{$\log{g}$}

\newcommand{\ki}{\ion{K}{1}}

\newcommand{\meth}{CH$_4$}
\newcommand{\water}{H$_2$O}
\newcommand{\wat}{H$_2$O}

\newcommand{\hh}{H$_2$}
\newcommand{\kms}{km~s$^{-1}$}
\newcommand{\name}{2MASS~J05325346+8246465}
\newcommand{\namesh}{2MASS~J0532+8246}

\slugcomment{Accepted to ApJ 19 July 2007; Accepted 10 September 2007}

\shorttitle{\namesh}
\shortauthors{Burgasser et al.}

\begin{document}

\title{Parallax and Luminosity Measurements of an L Subdwarf}

\author{Adam J.\ Burgasser\altaffilmark{1},
Frederick J.\ Vrba\altaffilmark{2}, 
S\'{e}bastien L\'{e}pine\altaffilmark{3},
Jeffrey A.\ Munn\altaffilmark{2},
Christian B.\ Luginbuhl\altaffilmark{2},
Arne A.\ Henden\altaffilmark{4},
Harry H.\ Guetter\altaffilmark{2},
and Blaise C.\ Canzian\altaffilmark{5}}

\altaffiltext{1}{Massachusetts Institute of Technology, Kavli Institute for Astrophysics and Space Research,
Building 37, Room 664B, 77 Massachusetts Avenue, Cambridge, MA 02139, USA; ajb@mit.edu}
\altaffiltext{2}{US Naval Observatory, Flagstaff Station, 10391 W. Naval Observatory Rd., Flagstaff, AZ 86001, USA}
\altaffiltext{3}{Department of Astrophysics, Division of Physical Sciences, American Museum of Natural History, Central Park West at 79th Street, New York, NY 10024, USA}
\altaffiltext{4}{AAVSO, 25 Birch St., Cambridge, 
MA, 02138-1205, USA}
\altaffiltext{5}{L-3 Communications/Brashear, 615 
Epsilon Dr., Pittsburgh, PA 15238, USA}

\begin{abstract}
We present the first parallax and luminosity 
measurements for an L subdwarf, the sdL7 {\name}.
Observations conducted over three years by the USNO 
infrared astrometry program yield an astrometric
distance of 26.7$\pm$1.2~pc
and a proper motion of 2.6241$\pm$0.0018$\arcsec$~yr$^{-1}$.
Combined with broadband spectral and photometric measurements,
we determine a luminosity of 
$\log{L_{bol}/L_{\sun}}$ = -4.24$\pm$0.06 and {\teff} = 1730$\pm$90~K
(the latter assuming an age of 5--10~Gyr), 
comparable to mid-type L field dwarfs. 
Comparison of the luminosity of {\name}
to theoretical evolutionary models indicates that its mass is
just below the sustained hydrogen burning limit, and is therefore
a brown dwarf.
Its kinematics 
indicate a $\sim$110~Myr, retrograde Galactic orbit which is both 
eccentric (3 $\lesssim$ R $\lesssim$ 8.5~kpc)
and extends well away from the plane ($\Delta$Z = $\pm$2~kpc),
consistent with membership in the inner halo population.
The relatively bright $J$-band magnitude of {\name} implies
significantly reduced opacity in the 1.2~$\micron$ region,
consistent with inhibited condensate formation as previously
proposed.  Its as yet unknown subsolar metallicity remains
the primary limitation in constraining its mass; determination
of both parameters
would provide a powerful test of interior and evolutionary models
for low-mass stars and brown dwarfs.
\end{abstract}


\section{Introduction}

The lowest luminosity stars and brown dwarfs 
are among the most useful probes of planetary, stellar and Galactic processes.
With hydrogen burning lifetimes far in excess of a Hubble time 
(e.g., \citealt{lau97}) and space densities
that exceed those of hotter stars, 
low-mass dwarfs are ubiquitous in the disk,
thick disk and halo populations (e.g., \citealt{dah95,rei02,dig03,cru07}),
and may host the bulk of terrestrial planets in the Galaxy (e.g., \citealt{bos06,tar07}).
The steady cooling of brown dwarfs over time 
makes them useful chronometers for coeval clusters \citep{bil97,sta98},
and these sources probe star formation
down to its lowest mass limit 
\citep[and references therein]{luh07}.  Observations
of cool brown dwarfs provide empirical constraints
on atmospheric chemical models, opacities and dynamics
(e.g., \citealt{ack01,lod02,hel04}),
and facilitate
studies of hot exoplanetary atmospheres (e.g., \citealt{bar03}).

Over the past decade, hundreds of low-mass stars and brown dwarfs 
have been identified by wide-field optical and near-infrared surveys 
such as the Two Micron All Sky Survey (2MASS; \citealt{skr06}), 
the Sloan Digital Sky Survey (SDSS; \citealt{yor00})
and the Deep Near Infrared Survey of the Southern Sky (DENIS; \citealt{epc97}).
These discoveries 
include members of two new spectral classes, the L dwarfs and T dwarfs \citep[and references therein]{kir05}\footnote{A current list of known L 
and T dwarfs is maintained at 
\url{http://dwarfarchives.org}.}.
As the number of low-luminosity dwarfs grows, 
distinct populations of ``peculiar'' sources are
being found which exhibit unusual surface gravities
(e.g., \citealt{kir06}), metallicities (e.g., \citealt{me0532,lep1610}),
or atmospheric structures (e.g., \citealt{cru03,cru07,kna04}).  
One such population
is the L subdwarf class \citep{mecs13}, the
metal-poor counterparts to solar-metallicity
field L dwarfs and the low-temperature extension of the M subdwarf
sequence \citep{giz97}. 
L subdwarfs are distinguished from L dwarfs
by the presence of relatively enhanced metal-hydride 
absorption bands (CaH, FeH, CrH) 
and metal lines (\ion{Ca}{1}, \ion{Ti}{1}, \ion{Fe}{1}) 
relative to reduced metal-oxide absorption (TiO, VO; \citealt{mou76}).
They also exhibit exceptionally blue near-infrared
colors ($J-K_s \lesssim 0$ compared to $J-K_s \approx 1.5-2.5$)
caused by strong collision-induced {\hh} absorption
\citep{lin69,sau94,bor97}.
Their halo or thick disk
kinematics \citep{me0532,rei06} indicate 
an origin early in the Galaxy's history.
L subdwarfs are useful for studying 
metallicity effects on cool atmospheric chemistry, 
particularly at temperatures in which photospheric condensates
first become an important source of opacity 
\citep{ack01,all01,bur06}.  Like L dwarfs, they may also span
the metal-dependent hydrogen-burning mass limit \citep{me0532},
and are therefore probes of low-mass star formation in the
metal-poor halo.  However, 
L subdwarfs are also exceptionally rare.  Only four
are currently known,
identified serendipitously in the 2MASS \citep{me0532,me1626}, 
SDSS \citep{siv07} and SUPERBLINK 
\citep{lep1610} surveys.

Understanding the physical characteristics of metal-poor
low-mass stars and brown dwarfs
requires the characterization of basic observational properties
--- distance, luminosity and effective temperature ({\teff}) ---
which can be 
facilitated by parallax measurements.  However, while
astrometric studies of late-type field dwarfs have provided 
robust absolute magnitude, {\teff} and luminosity scales 
down to the lowest luminosity brown dwarfs known
\citep{dah02,tin03,vrb04}, 
parallax measurements for late-type subdwarfs are rare \citep{mon92}, 
and there are no such measurements for any L subdwarfs.

To address this deficiency, we report the first parallax measurement
of an L subdwarf, 
{\name} (hereafter {\namesh}; \citealt{me0532}),
one of the first and latest-type L subdwarfs to be identified.
This source is tentatively classified
sdL7 due to the similarity of its optical spectrum to those of 
solar-metallicity L7 dwarfs
\citep{megmos}. Its late spectral type, low estimated temperature ({\teff} $\approx$ 1400--2000~K) and halo kinematics
all make {\namesh} a strong candidate halo brown dwarf. 

In $\S$~2 we describe astrometric and photometric measurements
of {\namesh}, conducted as part of the United States
Naval Observatory (USNO) infrared astrometry program \citep{vrb04},
and compare its absolute photometry to field late-type M, L and T dwarfs.
In $\S$~3 we calculate the bolometric luminosity and {\teff} of
{\namesh} based on our astrometric measurements and broad-band spectral
and photometric measurements reported in the literature.
In $\S$~4 we analyze these results, determining estimates of
the physical properties of {\namesh} using evolutionary models from \citet{bar97,bar98} and \citet{bur01}, and examine the kinematics of this source and its Galactic orbit.  Results are summarized in $\S$~5.

\section{Observations}

\subsection{Astrometric Measurements}

Astrometric observations of {\namesh} were obtained with 
the ASTROCAM near-infrared imager \citep{fis03}
at the USNO Flagstaff Station 61-inch Kaj Straand 
Astrometric Reflector
on 37 nights spanning a $\sim$3~yr 
period beginning in February 2003 and ending in 
February 2006.\footnote{Astrometric observations of 
{\namesh} were terminated only
after a June 2006 cryogenic explosion seriously damaged ASTROCAM. When
this instrument is restored to operational status (expected mid-2008), 
observations will resume for this source and other L and T dwarfs
in the USNO infrared astrometric program.}
Each of the observations was made in the $H$-band (1.7~$\micron$),
with three dithered exposures of 450 to 900~s each, depending
on seeing conditions (always less than 2$\farcs$5), 
or 1350 to 2700~s total integration per visit.
Data acquisition and astrometric reduction procedures
are discussed in detail in \citet{vrb04}.  Note that observations for
{\namesh} span a longer baseline than the L and T dwarfs reported
in \citet{vrb04}, so that stable parallaxes in both right ascension
and declination could be determined and combined. 
Twelve stars were employed in the reference frame,
and photometric parallaxes for these sources 
were determined using 2MASS photometry transformed to the CIT
system. The correction 
from relative to absolute parallax was found to be
1.14$\pm$0.13 mas.

Final values for the parallax (relative and absolute) and 
proper motion solutions for {\namesh}
are listed in Table~\ref{tab:usno}.
The astrometric distance of this source is 26.7$\pm$1.2~pc, within
the 10-30 pc range estimated by \citet{me0532}.
The proper motion measurement, 2.6241$\pm$0.0018 $\arcsec$ yr$^{-1}$, is also consistent with but substantially more accurate than
the previous determination (2.60$\pm$0.15 $\arcsec$ yr$^{-1}$).
The USNO measurements confirm the high tangential space velocity of {\namesh},
$V_{tan}$ = 332$\pm$15~{\kms}, over three
times larger than any of the known ``disk'' field L dwarfs 
\citep{vrb04,sch07}.
Combined with its radial velocity 
($V_{rad}$ = $-172{\pm}1$~{\kms}; \citealt{rei06}), we determine
local standard of rest (LSR) velocities of
$(U,V,W)_{LSR}$ = $(-70,-354,78)$ $\pm$ $(9,13,7)$~{\kms}, assuming
a LSR solar velocity of (10.00, 5.25, 7.17)~{\kms} \citep{deh98}.
The substantial negative $V_{LSR}$ velocity implies that {\namesh} is
orbiting in a retrograde motion about the Galactic center; i.e., 
in the opposite sense
of the Galactic disk 
(assuming that $V_{disk}$ = 220~{\kms} in galactocentric coordinates; 
\citealt{ker86}).  This 
would appear to rule out membership in the
thick disk population \citep{chi00}, so {\namesh}
is likely to be a halo low-mass object.
The Galactic orbit of this source is discussed further in $\S$~4.2.

\subsection{Photometric Measurements}

Near-infrared photometric measurements 
of {\namesh} were also obtained during the course
of the astrometric observations in an attempt to improve upon existing 
2MASS photometry (uncertainties of 
0.06, 0.09, and 0.15 mag in $J$-, $H$-, and
$K_s$-bands, respectively).  
$J$-, $H$- and $K$-band 
observations on the CIT system \citep{gue03} were obtained 
with ASTROCAM on 2003 October 12
(UT). Conditions were clear with $\sim$0.9$\arcsec$ seeing. 
Three dithered exposures were obtained in each 
filter with total integration times of 720, 720, and 1500~s,
respectively. 
The ASTROCAM field of view contained 9 stars
significantly 
brighter than {\namesh}, with mean 2MASS $J$, $H$, and $K_s$ 
uncertainties of 0.03, 0.04, and 
0.07 mag, respectively. 2MASS photometry for these sources
were converted to the CIT system using the transformations of \citet{car01},
which were then employed as ``local photometric standards'' 
on the ASTROCAM frames.
Aperture photometry of these stars and {\namesh} was then
carried out using DAOPHOT in
the IRAF\footnote{IRAF is distributed by the National Optical
Astronomy Observatories, which are operated by the Association of
Universities for Research in Astronomy, Inc., under cooperative
agreement with the National Science Foundation.} environment on each of the individual dithers, and the combined
instrumental magnitudes and colors were used to determine 
transformations to the
CIT standard system.  Final results for {\namesh} are listed in
Table~\ref{tab:usno}.  We note that along with
reduced errors, the USNO photometry of this source 
is brighter and somewhat bluer than the 2MASS photometry, albeit within
the 3$\sigma$ uncertainties of the latter.

\subsection{Examination of Absolute Magnitudes}

\begin{figure}
\epsscale{0.9}
\plotone{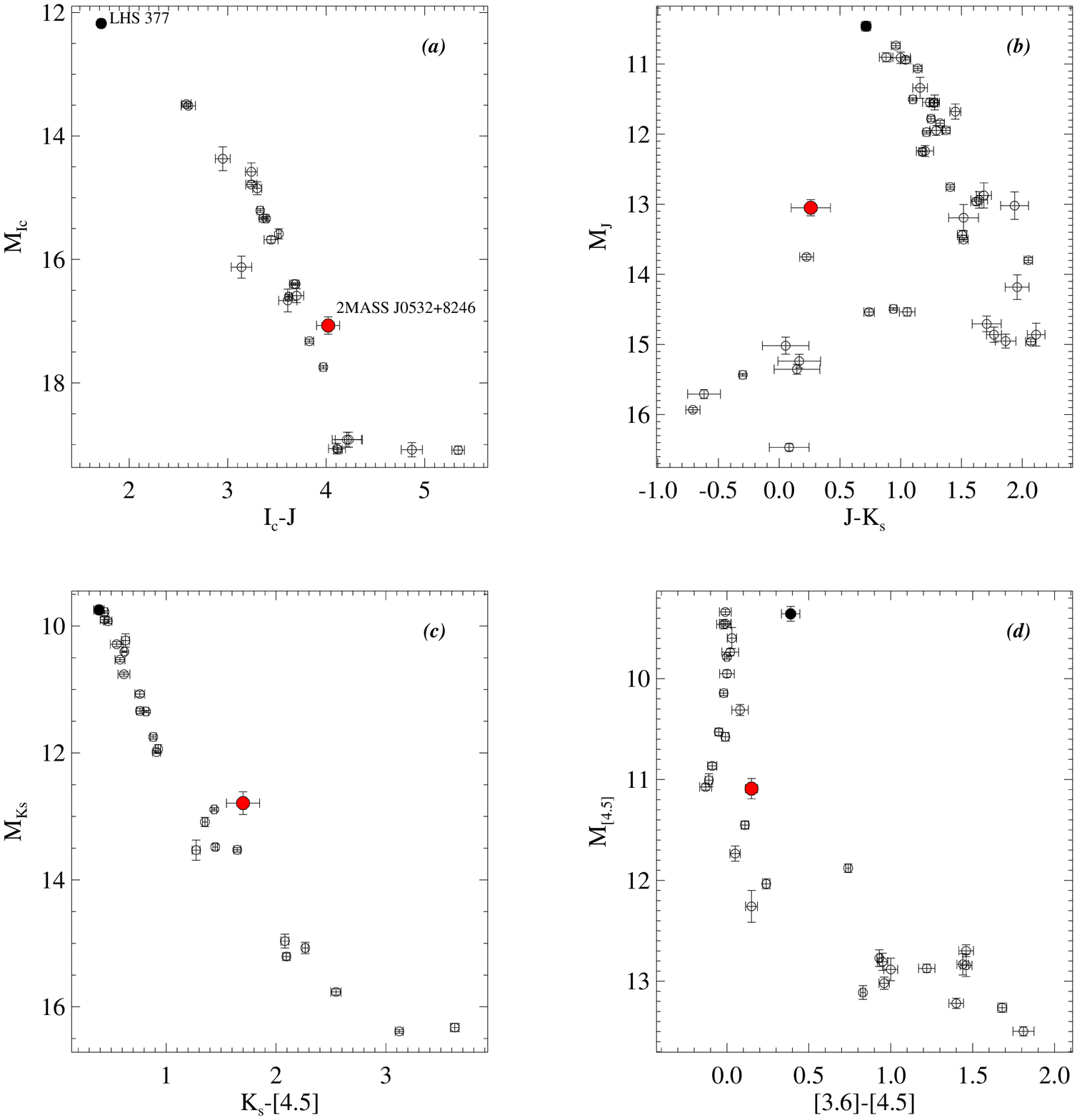}
\caption{Absolute magnitudes versus color for a sample of
field late-type M, L and T dwarfs, 
based on red optical and infrared photometry from 
\citet{mon92,dah02,me0532,pat06}; and 2MASS; 
and parallax measurements
from \citet{mon92,pry97,dah02,tin03}; and \citet{vrb04}.
Sources shown are constrained to have color and absolute magnitude uncertainties
$\leq$0.2 mag, and to be unresolved.
Diagrams shown are: 
(a) $M_{I_c}$ versus $(I_c-J)$, 
(b) $M_J$ versus $(J-K_s)$, 
(c) $M_{K_s}$ versus $(K_s-[4.5])$
and (d) $M_{[4.5]}$ versus $([3.6]-[4.5])$.
The location of {\namesh}
and the sdM7 LHS 377 are indicated in each panel by solid red 
and black circles, respectively. 
\label{fig:absmvscolor}}
\end{figure}

\begin{figure}
\epsscale{0.9}
\plotone{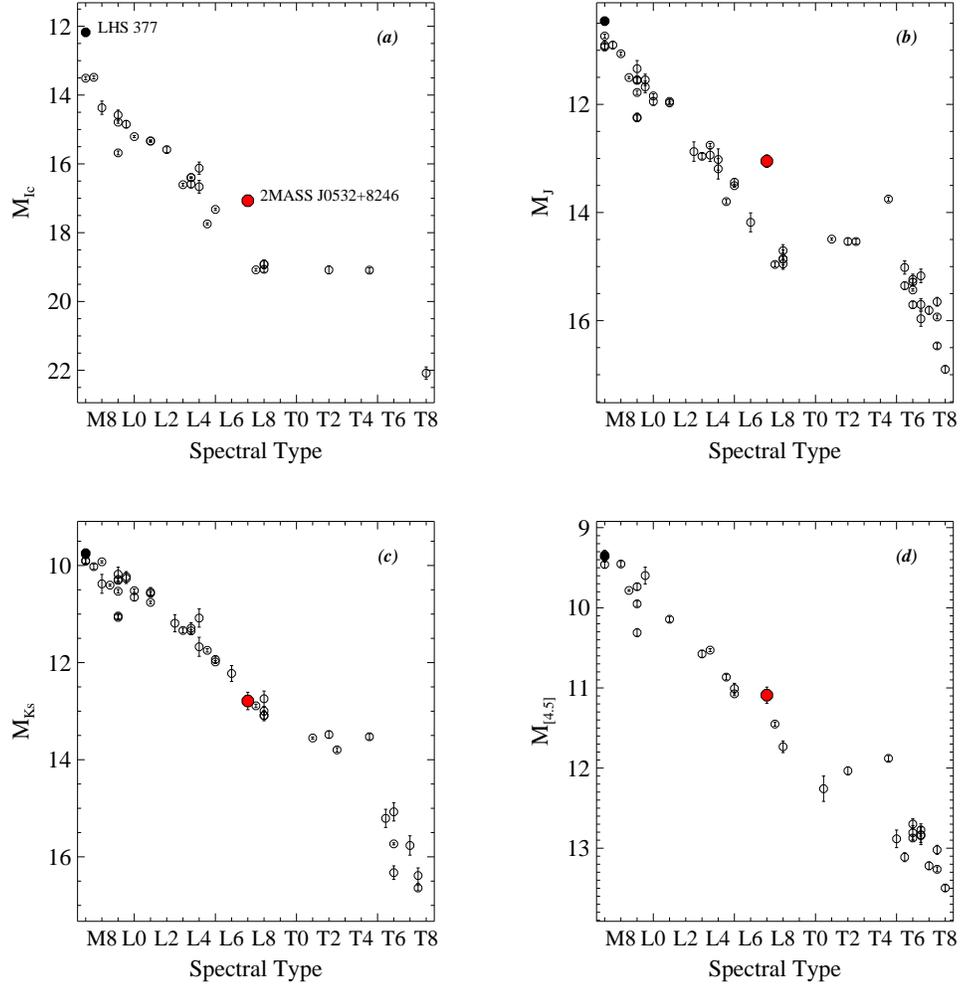}
\caption{Absolute magnitudes versus spectral type for the same sample
described in 
Figure~\ref{fig:absmvscolor}. 
Shown are (a) $M_{I_c}$, (b) $M_J$, (c) $M_{K_s}$ and (d) $M_{[4.5]}$
versus optical spectral types for M and L dwarfs \citep{kir99}
and near-infrared
spectral types for T dwarfs \citep{meclass2}.  Spectral
types of sdM7 and sdL7 are assumed for LHS 377 and
{\namesh} (solid black and red circles, respectively). 
\label{fig:absmvsspt}}
\end{figure}

{\namesh} is the first L subdwarf to have a measured distance;
hence, examination of its absolute magnitudes
(Table~\ref{tab:properties}) is of some importance in understanding
the atmospheric properties of low-temperature metal-poor dwarfs.
Figures~\ref{fig:absmvscolor}
and~\ref{fig:absmvsspt} display a sample of
optical and infrared color/magnitude
and spectral type/magnitude diagrams
for late-type M, L and T dwarfs with parallax measurements,
including {\namesh}.
Photometric data were culled from \citet{dah02} and \citet{me0532}
for $I_c$-band (0.8~$\micron$); 2MASS for near-infrared
$JHK_s$, and \citet{pat06} for mid-infrared measurements made
with the {\em Spitzer} IRAC instrument \citep{faz04}.
Astrometric data are from \citet{pry97,dah02,tin03};
and \citet{vrb04}.  For the spectral type/magnitude plots, we used
published optical spectral types for late-type M and L dwarfs (e.g., \citealt{kir99})
and near-infrared spectral types for T dwarfs (e.g., \citealt{meclass2}).
Sources were constrained to have color and magnitude uncertainties
no greater than 0.2~mag, and not known to be multiple
\citep[and references therein]{meppv}.
We also include absolute magnitudes and colors for the sdM7 LHS~377 \citep{mon92,giz97}, until now the latest-type subdwarf with a 
reported parallax measurement.

In general, {\namesh} is somewhat overluminous for both its color
and spectral type.  This is 
particularly the case at $J$-band, where it
is 1--2 mag brighter than the disk L dwarf/T dwarf sequence
based on both spectral type and color.
Indeed, {\namesh} lies in a relatively unpopulated region
in the $M_J$ versus $(J-K_s)$ color/magnitude diagram.
$J$-band opacity in L dwarf photospheres is dominated by
absorption from condensates, which
gives rise to their red $J-K_s$ colors
\citep{tsu96,tsu99,bur99,ack01,all01}.
\citet{me0532}, \citet{rei06} and \citet{megmos} have all
speculated that condensate formation may be inhibited in 
L subdwarf photospheres
based on the unexpected strength of gaseous TiO, \ion{Ca}{1} and \ion{Ti}{1}
features.
Reduced condensate opacity 
allows $J$-band light to escape from deeper and hotter
layers, resulting in an overall brightening at these wavelengths
\citep{ack01}.  
In contrast, $K$-band opacity is dominated by 
collision-induced {\hh} opacity in late-type dwarfs, a species whose
absolute abundance is not modified by metallicity. 
The higher pressure photospheres of old, high surface gravity 
subdwarfs will in fact increase
{\hh} opacity.  This trend explains why the $M_{K_s}$ magnitude of {\namesh} is
consistent with, and perhaps slightly fainter than, those of 
solar-metallicity L7 field dwarfs.
Given that the primary deviation in the near-infrared brightness
of {\namesh} occurs in the spectral region in which 
condensate opacity 
should be dominant, these results provide further support for the
idea that condensate formation is inhibited in low-temperature
metal-poor atmospheres.

The location
of {\namesh} redward and/or above the L dwarf sequence in the 
$M_{I_c}$ versus $(I_c-J)$
and $M_{K_s}$ versus $(K_s-[4.5])$ color/magnitude diagrams
can also be explained by metallicity effects.  
Reduced $J$-band condensate 
opacity coupled with residual (albeit reduced) 
TiO and pressure-broadened {\ki} absorption
at $I_c$-band result in slightly redder $(I_c-J)$ colors
than comparably classified field dwarfs.  
This is in contrast to the sdM7 LHS~377 which is brighter at $I_c$-band 
rather than $J$-band as compared to M7 field dwarfs, and hence bluer 
in $(I_c-J)$.  
The difference in color peculiarity between these two sources 
can be attributed to the fact that
late M dwarfs are too warm
to have significant condensates in their photospheres \citep{ack01},
while variations in metal oxide absorption have a greater effect.
The $(K_s-[4.5])$ color of {\namesh}
is slightly redder than field dwarfs with similar $M_{K_s}$ magnitudes
because of enhanced 4.5~$\micron$ flux.  This region is 
dominated by metallicity-sensitive molecular CO and {\water} opacity, 
in contrast to the 
to strong {\hh} absorption at $K_s$.  
Both {\namesh} and LHS~377 have redder $([3.6]-[4.5])$ colors
than the dwarf sequence; similar metallicity-induced color 
effects have also been noted amongst field T dwarfs \citep{lie07,leg07}.\footnote{Late-type esdMs also have redder $B-V$ colors at a given
$M_V$ magnitude due to similar 
metal opacity effects \citep{giz97}.}
The red $([3.6]-[4.5])$ color of {\namesh} may be muted, however,
if {\meth} absorption at 3.3~$\micron$, present in mid- and late-type
L dwarfs \citep{nol00,cus05} is also weakened by metallicity effects.

\section{Bolometric Luminosity and Effective Temperature Determinations}

\begin{figure}
\epsscale{0.9}
\plotone{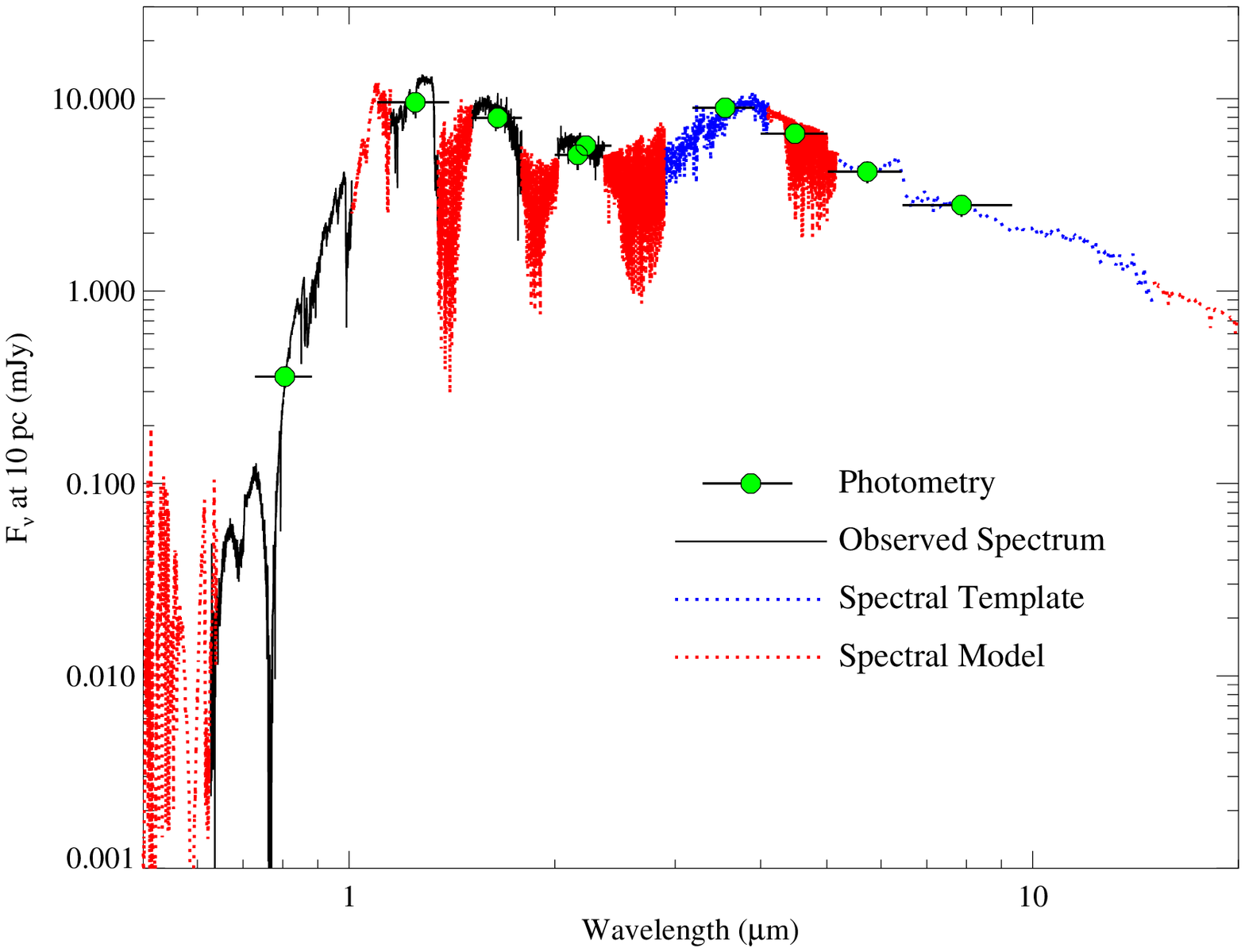}
\caption{Absolute spectral energy distribution (F$_{\nu}$ at 10~pc) 
of {\namesh} used to calculate its bolometric luminosity.
Absolute photometry from \citet{me0532}; 2MASS; \citet{pat06};
and this paper are indicated by solid green dots.  Flux-calibrated spectral
data from \citet{me0532} are indicated by black lines.  Scaled spectral template
data for the L5 2MASS~J1507-1627 
\citep{cus06} are indicated by blue lines.  Scaled NEXTGEN
spectral model data ({\teff} = 2000~K, {\logg} = 5.5 cgs
and [M/H] = -1.0) from \citet{hau99} are indicated by red lines.
This hybrid spectrum was integrated to determine a bolometric flux
of (18.2$\pm$2.6)$\times$10$^{-12}$ erg~cm$^{-2}$~s$^{-1}$, which includes an
estimate of systematic uncertainty.
\label{fig:sed}}
\end{figure}

Our distance measurement for {\namesh} enables calculation
of its bolometric luminosity and
{\teff}, through the use of existing spectroscopic and photometric
data spanning 0.6--15~$\micron$.  Our baseline calculation,
illustrated in Figure~\ref{fig:sed}, was constructed as follows.
First, near-infrared spectral data for
{\namesh} from \citet{me0532} were piece-wise scaled 
to absolute 2MASS $J$, $H$ and USNO $K$-band magnitudes
(Table~\ref{tab:properties}). 
Gaps in the {\wat} bands and the 1~$\micron$ peak were
substituted by a NEXTGEN spectral model \citep{hau99}
with parameters {\teff} = 2000~K, 
{\logg} = 5.5, [M/H] = -1.0 dex, scaled to match the 
flux-calibrated spectral data in the near-infrared peaks.
Red optical spectral data for {\namesh}
were scaled to the absolute $I_c$ magnitude of this source \citep{me0532}.  
As there are no reported spectral data for {\namesh}
at longer wavelengths,
we used a spectral template based on data 
for the L5 dwarf
2MASS~J15074769-1627386 \citep[hereafter 2MASS~J1507-1627]{rei00,cus06},
spanning the 2.9--4.1 and 5.2--15.4~$\micron$ regions and piece-wise
scaled to match the absolute 3.6, 5.8 and 8.0~$\micron$ {\em Spitzer} IRAC
photometry of {\namesh} \citep{pat06}.  Flux calibration of these
data was done using
the appropriate correction factor for IRAC photometry
as discussed in \citet{cus06}.
The 2.4--2.9 and 4.1--5.2~$\micron$ gaps were again 
replaced with theoretical model spectra, scaled to overlap with the template 
spectra and to match absolute 4.5~$\micron$ photometry of {\namesh}.
Short (0.1 $<$ $\lambda < 0.6$~{\micron}) and long (15 $<$ $\lambda < 1000$~{\micron}) 
wavelength regions were calculated by scaling the NEXTGEN
spectral model to the observed and template spectra,
respectively.  

The resulting broad spectral energy distribution was
then integrated to determine the total spectral flux, with nominal
uncertainties based on the absolute photometry, including 
astrometric uncertainties.
To examine systematic effects, we repeated our analysis
replacing the NEXTGEN model with a {\teff} = 1700~K, 
{\logg} = 5.5., [M/H] = -0.5
condensate cloud model from \citet[hereafter, Tucson model]{bur06},
as well as linearly interpolating over gaps in the observed
and template spectral data.

Table~\ref{tab:sed} provides a breakdown of the total fluxes 
of {\namesh} in five
spectral regions, with comparisons between the different 
computational methods.  As expected, the bulk of the spectral flux
for this late-type dwarf arises at infrared wavelengths ($\sim$93\%),
with the 1.0--2.9~$\micron$ region encompassing almost 70\% of the
total light.  Differences between the computational
methods are
generally less than the formal uncertainties, except 
in the shortest and longest wavelength regions which fortuitously contribute
negligibly to the aggregate flux ($<$0.25\%).  Bolometric flux values
computed using the two spectral models and linear interpolation
are within 3$\sigma$ of each other,
with our baseline calculation providing the median value.
We therefore use this value as the measured bolometic flux, and
propagate the estimated systematic error ($\sim$14\%)
into our final uncertainties.

We determine an absolute bolometric flux of  
(18.2$\pm$2.6)$\times$10$^{-12}$ erg~cm$^{-2}$~s$^{-1}$
for {\namesh}, which translates
into a bolometric luminosity of $\log{L/L_{bol}}$ = -4.24$\pm$0.06,
or $M_{bol}$ = 15.35$\pm$0.16.  This is comparable to mid-type L dwarfs
like 2MASS~J1507-1627 ($M_{bol}$ = 15.41$\pm$0.13; \citealt{vrb04}),
and is roughly half a magnitude more luminous than the average L7 dwarf.
{\namesh} is nevertheless over 10 times less luminous than the 
next coolest subdwarf with a parallax measurement,
sdM7 LHS 377 ($\log{L/L_{bol}}$ = -3.11$\pm$0.02; \citealt{leg00}).
Bolometric corrections ($BC_b \equiv M_{bol} - M_b$)
at $J$ and $K$ were determined
as $BC_J$ = 2.30$\pm$0.08 and $BC_K$ = 2.68$\pm$0.09. 
The latter is $\sim$0.5 mag 
smaller than comparable values computed for L6--L8 field dwarfs,
further illustrating the substantial redistribution of flux
from this metal-poor source.

An effective temperature for {\namesh} can be estimated 
assuming that
this source is likely older than $\sim$5~Gyr based on its
kinematics, and has a mass near the hydrogen burning
limit (0.05--0.09~M$_{\sun}$, depending on metallicity; see $\S$~4.1).
These assumptions imply a radius
of 0.096$\pm$0.015~R$_{\sun}$ based on the solar-metallicity
evolutionary  models of \citet{bur01} and \citet{bar03},
and yields {\teff} = 1600$\pm$300~K, where the uncertainty is entirely
dominated by the uncertainty in the radius.   If we include
the measured luminosity (including 3$\sigma$ uncertainty)
as an additional constraint on these models,
a more refined radius estimate of 0.084$\pm$0.003~R$_{\sun}$
is obtained, corresponding to {\teff} = 1730$\pm$90~K.  Again, this 
is comparable to temperatures for mid-type L dwarfs \citep{vrb04,gol04}
and is nearly 1200~K cooler than LHS~377 \citep{leg00}

\section{Discussion}

\subsection{The Substellar Nature of {\namesh}}

\begin{figure}
\epsscale{1.1}
\plottwo{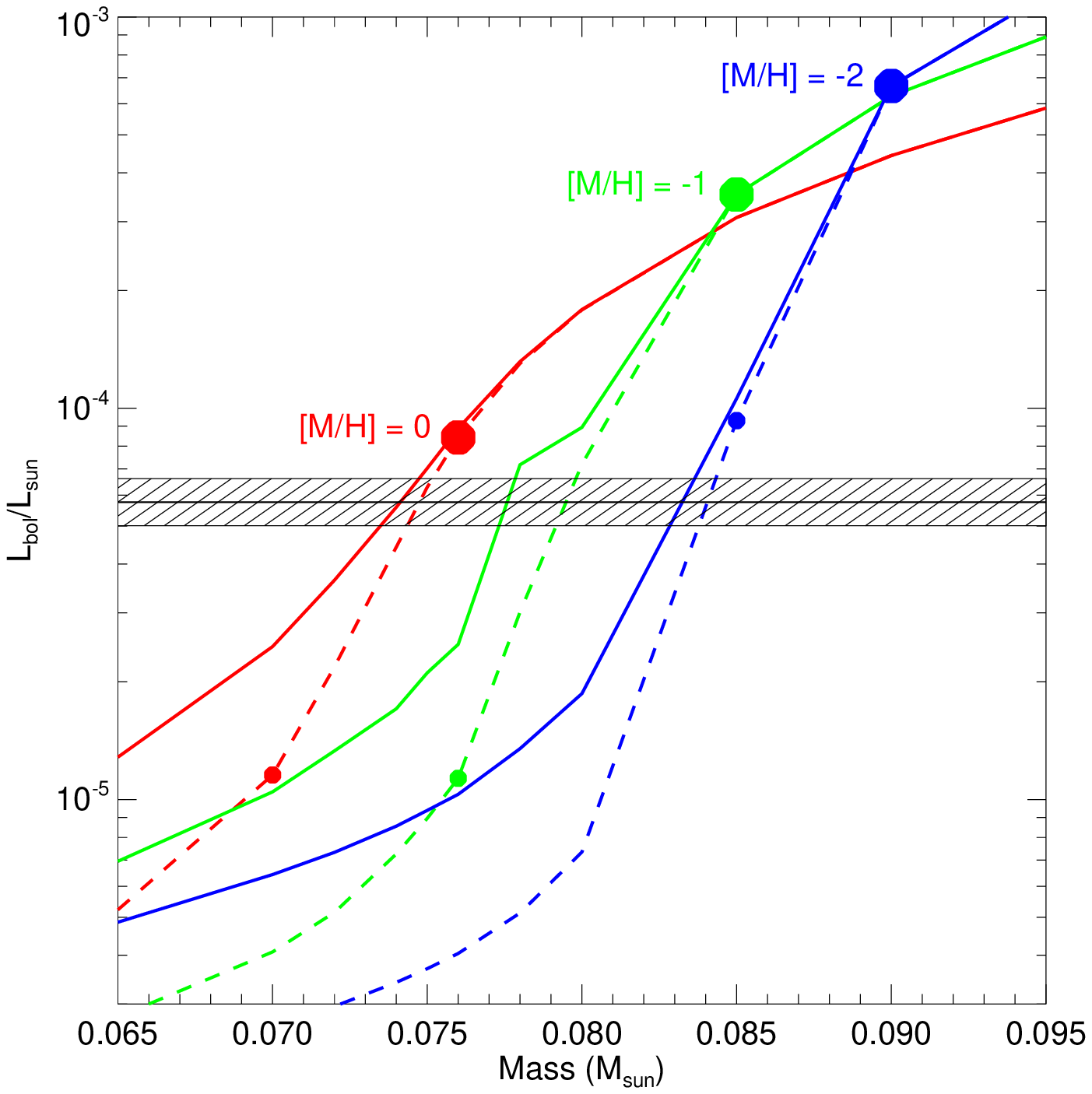}{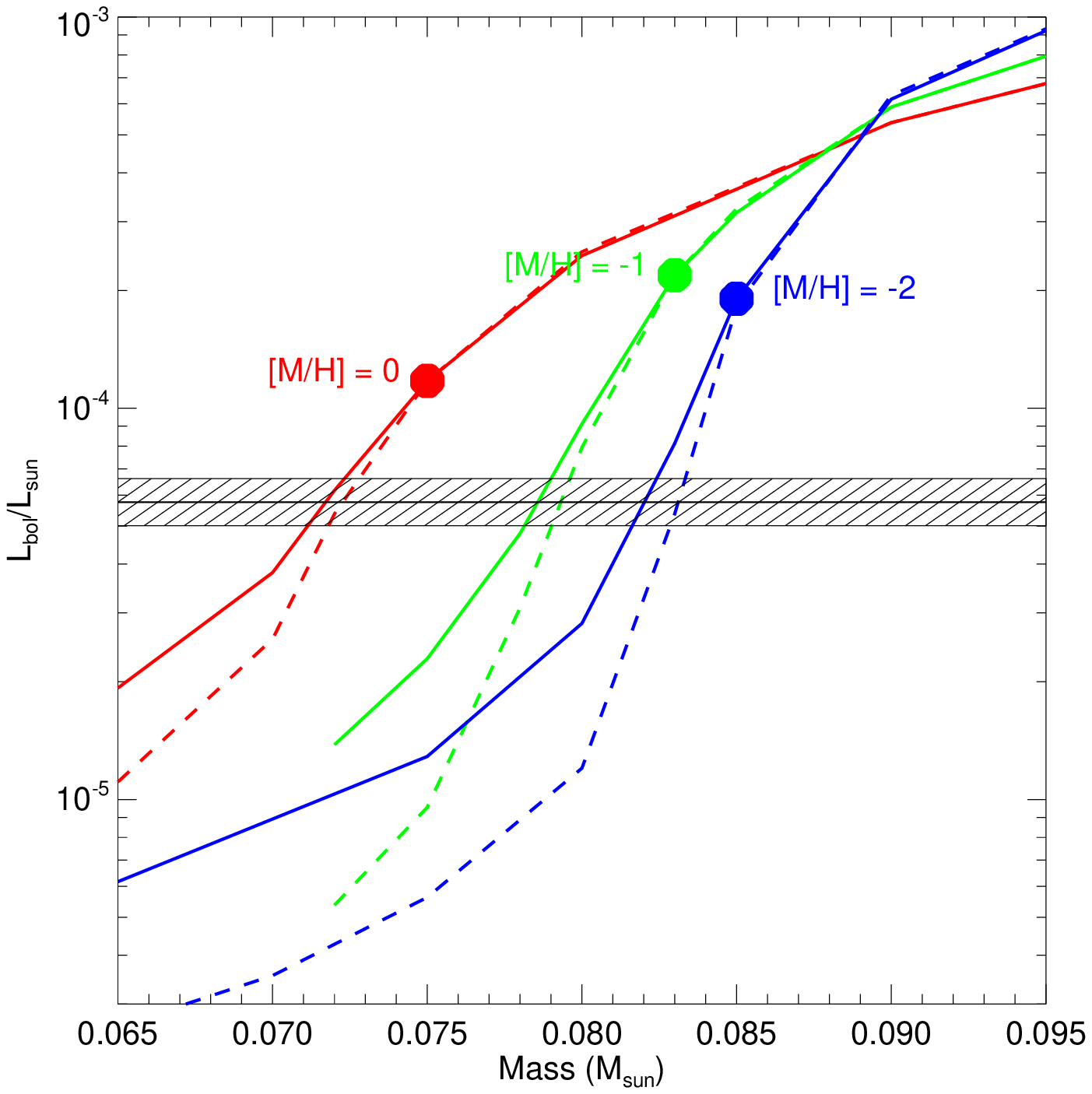}
\caption{Theoretical mass/luminosity relations for low-mass stars
and brown dwarfs, based on evolutionary models from \citet[left panel]{bur01}
and \citet[right panel]{bar97,bar98}.
Models are shown for ages of 5 (solid lines) and 10~Gyr (dashed lines) 
and metallicities [M/H] = 0 (red lines), -1 (green lines) 
and -2 dex (blue lines).  The limits at which 50\% (small circles in left panel)
and 99\% (large circles in both panels) of the total luminosity originates from
core hydrogen fusion are indicated.  The measured luminosity of
{\namesh} is indicated by the hashed region (1~$\sigma$ uncertainties),
and falls below the 99\% limit for all models, but above
the 50\% limit for [M/H] $\gtrsim$ -1 models of \citet{bur01}.  
Note that each model predicts a slightly different mass
for {\namesh}, varying primarily by metallicity but according
to evolutionary calculation as well.
\label{fig:evol}}
\end{figure}

While parallax and bolometric luminosity measurements
are able to constrain some of the atmospheric properties of
{\namesh}, they do not directly address the question of whether
this source is substellar.  For this we require comparison 
to evolutionary models; in particular, models incorporating subsolar
metallicities.  The reduced opacity of a metal-poor
atmosphere results in more rapid cooling;
hence, the luminosity of a metal-poor brown dwarf will be lower
than that of a solar metallicity brown dwarf
with the same mass and age, while the hydrogen burning minimum mass limit 
increases for lower metallicities (e.g., \citealt{dan85}). 
Figure~\ref{fig:evol} illustrates these trends by
comparing theoretical mass/luminosity relations from \citet{bur01}
and \citet{bar97,bar98},
spanning the hydrogen burning limit
for ages of 5 and 10~Gyr and metallicities [M/H] = 0, -1 and -2 dex.
The limits at which 50\% and 99\% of the luminosity is generated
from core hydrogen fusion at 10~Gyr are indicated; note how
these limits shift to higher masses and luminosities for
lower metallicity models.  The luminosity of {\namesh}
falls below the 99\% limits for all models
shown, as well as for intermediate and lower metallicities.
This appears to confirm {\namesh} as a brown dwarf, 
incapable of sustaining its luminosity by core hydrogen 
fusion alone.  However, note that for
[M/H] $\gtrsim$ -1, hydrogen fusion 
provides over half of the energy emitted, so {\namesh}
is fairly close to the stellar/substellar boundary.  Indeed, 
if its metallicity is sufficiently high, 
this object may eventually cool to the point at which its
lower luminosity is sustained by core fusion,
changing this low-mass brown dwarf into a star.

The luminosity of {\namesh} falls on a particularly steep 
section of the mass/luminosity relations shown,
enabling relatively tight (1\%), albeit model-dependent, constraints on
the mass of this source for a given metallicity.  
However, the variation in the derived mass for the
metallicity range shown is considerably larger, ranging from 0.0744 to
0.0835~M$_{\sun}$ for [M/H] = 0 to -2 for the \citet{bur01}
models (these values are consistent with mass estimates
based on the \citet{bar97,bar98} models).  Our best guess
for the metallicity of this source, [M/H] = -1 (\citealt{megmos};
see also \citealt{sch1444}), implies a mass
of 0.0783$\pm$0.0013~M$_{\sun}$ (0.0788$\pm$0.0006~M$_{\sun}$
for the \citealt{bar97,bar98} models). However, 
it is clear that the
unknown metallicity of {\namesh} remains the largest 
source of uncertainty in characterizing
its physical properties.

\subsection{The Galactic Orbit of {\namesh}}

\begin{figure}
\epsscale{0.9}
\plotone{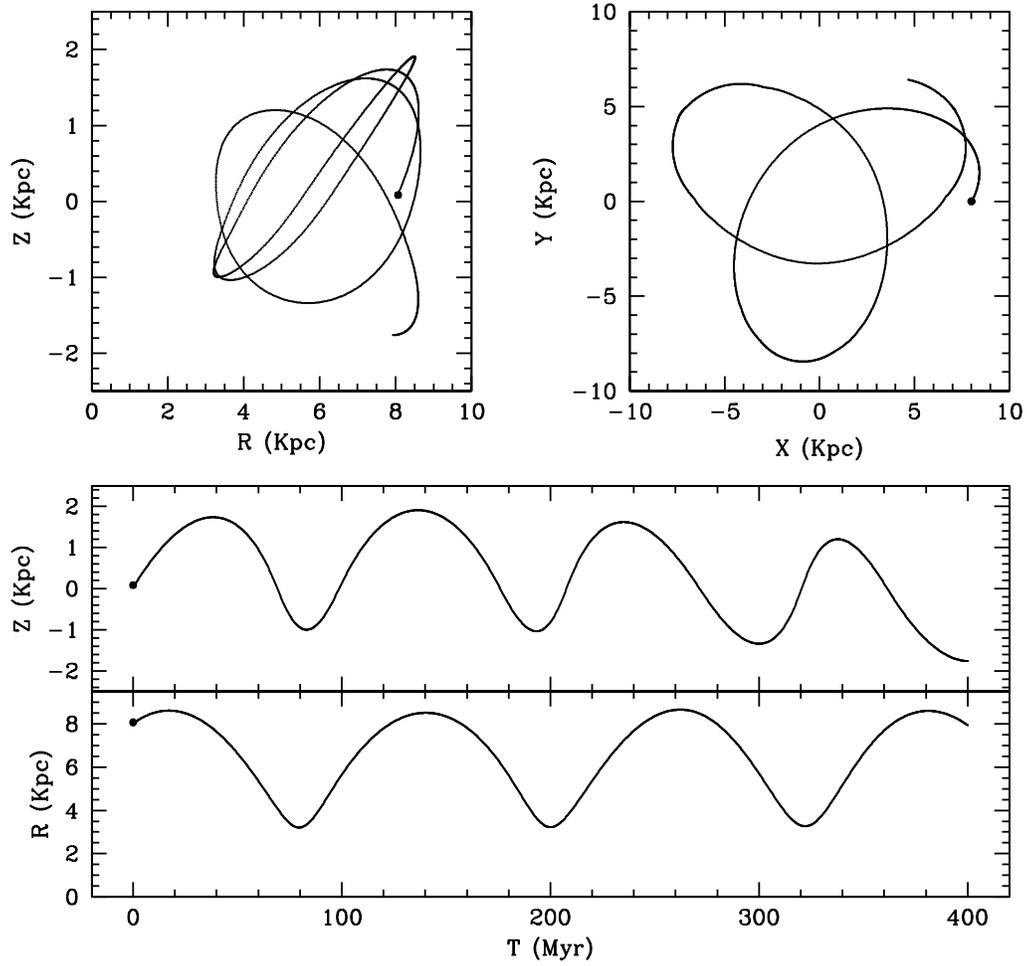}
\caption{Integrated Galactic orbit of {\namesh}, based on its $(U,V,W)_{LSR}$
space velocities, 2MASS coordinates and measured distance.  The orbit was
calculated using the Galactic mass model of \citet{dau95}
and a Runge-Kutta fourth-order integrator.
The top left panel shows the projected orbit perpendicular
to the Galactic disk, with the Galactic plane at Z = 0.
The top right panel shows the projected orbit in the plane of the
Galaxy; the sense of the Galactic disk rotation is clockwise.
The bottom panels show the temporal evolution of the orbit in 
Galactocentric cylindrical coordinates over a period of 400~Myr
starting from the current epoch (roughly four revolutions).
\label{fig:orbit}}
\end{figure}

The kinematics of {\namesh} argue strongly for membership in the
Galactic halo, but was this source formed early in the Galaxy 
or was it tidally stripped from one of the Galaxy's dwarf satellites
during a merging event?
To examine this question, we calculated
a probable orbit for {\namesh} based on its measured kinematics,
distance and 2MASS coordinates (epoch 1999 March 1 UT; Table~\ref{tab:properties}).
We used the Galactic mass model of \citet{dau95}
and a Runge-Kutta fourth-order integrator; see \citet{lep1826} for details.
Figure~\ref{fig:orbit} displays the results of this calculation,
projecting the $\sim$110~Myr 
Galactocentric orbit in cylindrical coordinates,
as well as the evolution of Galactic radius and vertical scaleheight
over time.
{\namesh} appears to have a relatively eccentric orbit,
moving between roughly 3 and 8.5~kpc of the Galactic center ($e$ $\approx$ 0.5).
The vertical scaleheight of its orbit
extends $\sim$2~kpc above and below the Galactic plane.

The orbital characteristics of {\namesh} appear to be inconsistent
with tidal capture from an external system, as this source generally
remains interior to the Solar Galactic orbit.  Rather, it
is likely to be a member of the Galaxy's inner flattened halo \citep{som90},
indicating formation from the early
building blocks of the Galaxy (e.g., \citealt{nor94,car96}).
The substantial retrograde motion of {\namesh} 
also argues against formation and 
(violent) ejection from the inner disk.
In other words, {\namesh} is likely to be quite old.  However,
this does not provide a robust constraint on its metallicity.
Mean eccentricity/metallicity relations by \citet{car96} and
\citet{chi00} suggest [Fe/H] $\lesssim$ -1, but the latter
study also finds no clear correlation between orbital eccentricity 
and metallicity for [Fe/H] $<$ -0.8.  The chemical abundances present
in the atmosphere of {\namesh}, a necessary constraint for mass estimation,
must be determined through other means, most likely spectral modeling.

\section{Summary}

We have presented the first parallax and luminosity measurements
for an L subdwarf, the sdL7 {\namesh}.  The derived parameters, 
summarized in Table~\ref{tab:properties}, confirm the low-luminosity,
low-temperature nature of this source, similar in both respects
to mid-type field L dwarfs.  Examination of absolute photometry
indicates that this source is substantially brighter at $J$-band
than comparable solar-metallicity field dwarfs, consistent
with reduced condensate opacity as previously suggested by optical spectroscopy.
The low luminosity of {\namesh}, $\log{L_{bol}/L_{\sun}}$ = -4.24$\pm$0.06,
compared to theoretical evolutionary models also confirms
the substellar nature of this source (M $\approx$ 0.078~M$_{\sun}$,
assuming [M/H] = -1), although it is probably still
fusing hydrogen at a low level in its core.  
Its kinematics
are consistent with membership in the inner halo population, making
this source the first {\em bona-fide} halo brown dwarf.

Further investigation of {\namesh} should be directed toward determining
its metallicity and chemical abundances.  As discussed in $\S$~4.1, 
uncertainty in the estimated mass of this source is dominated by 
its unknown metallicity.
High resolution spectroscopy (e.g., \citealt{rei06}) and accurate
spectral modeling can improve mass constraints, 
although such analysis must also provide a consistent luminosity 
determination (cf., \citealt{smi03}).  Independent determination of both
metallicity and mass could also provide an important empirical test on the
evolutionary models themselves, and thus one of the few constraints on
interior brown dwarf physics (in addition to radii measurements; 
e.g., \citealt{sta06}).  
Studies of metallicity effects in the physical
and observational properties of low-mass stars and brown dwarfs
will also benefit from the measurement of
parallaxes for additional late-type subdwarfs, to fill the gap
between LHS~377 (sdM7) and {\namesh} (sdL7).  Such observations
should be of high priority, as detailed studies of metallicity
effects in the atmospheric properties and evolution of the
lowest-luminosity stars and brown dwarfs are currently limited by
the absence of these data.

\acknowledgements
The authors would like to thank I.\ Baraffe and A.\ Burrows for making
available electronic versions of their evolutionary models for our
analysis, and M.\ Cushing for making available his spectral 
data for 2MASS~J1507-1627..  We also thank J.\ Bochanski for additional 
comments to the manuscript and J.\ Gizis for his prompt review.
This publication makes
use of data from the Two Micron All Sky Survey, which is a joint
project of the University of Massachusetts and the Infrared
Processing and Analysis Center, and funded by the National
Aeronautics and Space Administration and the National Science
Foundation. 2MASS data were obtained from the NASA/IPAC Infrared
Science Archive, which is operated by the Jet Propulsion
Laboratory, California Institute of Technology, under contract
with the National Aeronautics and Space Administration.
This research has benefitted from the M, L, and T dwarf compendium housed at DwarfArchives.org and maintained by Chris Gelino, Davy Kirkpatrick, and Adam Burgasser;
and the VLM Binary Archive maintained by N. Siegler at 
\url{http://paperclip.as.arizona.edu/$\sim$nsiegler/VLM\_binaries/}.

Facilities: \facility{USNO(ASTROCAM)}

\clearpage

\begin{deluxetable}{ll}
\tabletypesize{\footnotesize}
\tablecaption{USNO Astrometric and Photometric Measurements for {\namesh}. \label{tab:usno}}
\tablewidth{0pt}
\tablehead{
\colhead{Parameter} &
\colhead{Value} \\
}
\startdata
$\pi_{rel}$  & 36.3$\pm$1.6 mas \\
$\pi_{abs}$& 37.5$\pm$1.7 mas \\
$\mu$  & 2.6241$\pm$0.0018$\arcsec$ yr$^{-1}$ \\
$\theta$  & 128.91$\degr\pm$0.02$\degr$ \\
$V_{\tan}$  & 332$\pm$15 {\kms} \\
$(J-H)$\tablenotemark{a}  & 0.121$\pm$0.017 mag \\
$(J-K)$\tablenotemark{a}  & 0.17$\pm$0.07 mag \\
$K$\tablenotemark{a}  & 14.80$\pm$0.07 mag \\
\enddata
\tablenotetext{a}{Magnitudes on the CIT photometric
system.}
\end{deluxetable}

\clearpage
\thispagestyle{empty}
\begin{deluxetable}{lccl}
\tabletypesize{\footnotesize}
\tablecaption{Integrated Spectral Flux for {\namesh}. \label{tab:sed}}
\tablewidth{0pt}
\tablehead{
\colhead{Wavelength Range} &
\colhead{Flux} &
\colhead{Fraction} &
\colhead{Method} \\
\colhead{($\micron$)} &
\colhead{(10$^{-12}$ erg cm$^{-2}$ s$^{-1}$)} &
\colhead{(\%)}  & \\
}
\startdata
0.01--0.64~$\micron$  & {\bf 0.031$\pm$0.003} & 0.17 & NEXTGEN Model, 2MASS photometry  \\
 & 0.096$\pm$0.009 & \nodata & Linear extrapolation, 2MASS photometry  \\

\noalign{\smallskip}

0.64--1.0~$\micron$ & {\bf 1.24$\pm$0.11} & 6.81 &  Observed spectrum, 2MASS photometry \\

\noalign{\smallskip}

1.0-2.9~$\micron$ & {\bf 12.6$\pm$0.5} & 69.45 &  Observed spectrum, NEXTGEN model, 2MASS/USNO photometry \\
 & 12.4$\pm$0.4 & \nodata &  Observed spectrum, Tucson model, 2MASS/USNO photometry \\
 & 12.7$\pm$0.4 & \nodata &  Observed spectrum, spectral template, linear interpolation, 2MASS/USNO photometry \\

\noalign{\smallskip}

2.9--15~$\micron$ & {\bf 4.27$\pm$0.19} & 23.52 &  Spectral template, NEXTGEN model, IRAC photometry \\
 & 4.4$\pm$0.5 & \nodata &  Spectral template, Tucson model, IRAC photometry \\
 & 4.39$\pm$0.12 & \nodata &  Spectral template, linear interpolation, IRAC photometry \\

\noalign{\smallskip}

15--1000~$\micron$ & {\bf 0.0096$\pm$0.0009} & 0.05 &  NEXTGEN model, IRAC photometry \\
 & 0.077$\pm$0.007 & \nodata &  Raleigh-Jeans tail, IRAC photometry \\

\noalign{\smallskip}
\cline{1-4}
\noalign{\smallskip}

0.1--1000~$\micron$  & {\bf 18.2$\pm$1.7} & 100.00 &  Observations + NEXTGEN model \\
 & 18.0$\pm$1.8 & \nodata &  Observations + Tucson model \\
 & 18.5$\pm$1.7 & \nodata &  Observations + linear interpolation \\
\enddata
\tablecomments{Values listed in bold represent our baseline calculation.}
\end{deluxetable}
\clearpage

\begin{deluxetable}{lll}
\tabletypesize{\footnotesize}
\tablecaption{Properties of {\namesh}. \label{tab:properties}}
\tablewidth{0pt}
\tablehead{
\colhead{Parameter} &
\colhead{Value}  &
\colhead{Ref}  \\
}
\startdata
$\alpha$\tablenotemark{a} & 05$^h$32$^m$53$\fs$46 & 1 \\
$\delta$\tablenotemark{a} & +82$\degr$46$\arcmin$46$\farcs$5 & 1 \\
$\mu$  & 2.6241$\pm$0.0018$\arcsec$ yr$^{-1}$ & 2 \\
$\theta$  & 128.91$\degr\pm$0.02$\degr$ & 2  \\
Spectral Type & sdL7 & 3,4 \\
$d$ & 26.7$\pm$1.2 pc &   2 \\
$M-m$ & $-2.13{\pm}0.10$ mag &  2 \\
$(U,V,W)_{LSR}$  & $(-70,-354,78){\pm}(9,13,7)$ {\kms} &  2,5 \\
Kinematic Pop. & Halo & 2 \\
$\log{L_{bol}/L_{\sun}}$ & $-4.24{\pm}0.06$ &  1 \\
$M_{bol}$  & 15.35$\pm$0.16 mag &  2 \\
$M_{I_c}$  & 17.07$\pm$0.14 mag &  2,3 \\
$M_J$  & 13.05$\pm$0.12 mag &  1,2 \\
$M_H$  & 12.77$\pm$0.13 mag &  1,2 \\
$M_{K_s}$  & 12.79$\pm$0.18 mag &  1,2 \\
$M_K$  & 12.67$\pm$0.12 mag &  2 \\
$M_{[3.6]}$  & 11.24$\pm$0.10 mag &  2,6 \\
$M_{[4.5]}$  & 11.09$\pm$0.10 mag &  2,6 \\
$M_{[5.8]}$  & 11.10$\pm$0.14 mag &  2,6 \\
$M_{[8.0]}$  & 10.90$\pm$0.14 mag &  2,6 \\
$BC_J$ & 2.30$\pm$0.08 mag & 1,2 \\
$BC_K$ & 2.68$\pm$0.09 mag & 2 \\
{\teff} & 1730$\pm$90 K & 2,7,8 \\
Mass\tablenotemark{b}  & 0.0744$\pm$0.0009 M$_{\sun}$ &  2,7 \\
 & 0.0783$\pm$0.0013 M$_{\sun}$ &  2,7 \\
 & 0.0825$\pm$0.0008 M$_{\sun}$ & 2,7 \\
\enddata
\tablenotetext{a}{J2000 coordinates at epoch 1999 March 1 (UT).}
\tablenotetext{b}{Mass estimates for [M/H] = 0, -1 and -2 based on
the evolutionary models of \citet{bur01}, assuming an age of 5--10~Gyr.
Values are consistent with those derived using the \citet{bar97,bar98} models,
within the estimated uncertainties.}
\tablerefs{(1) 2MASS; (2) This paper; (3) \citet{me0532}; 
(4) \citet{megmos}; (5) \citet{rei06}; 
(6) \citet{pat06}; (7) \citet{bur01}; (8) \citet{bar03}.}
\end{deluxetable}

\clearpage

\begin{thebibliography}{}

\bibitem[Ackerman \& Marley(2001)]{ack01}Ackerman, A.\ S., \& Marley, M.\ S.
2001, \apj, 556, 872

\bibitem[Allard et al.(2001)]{all01}Allard, F., Hauschildt, P.\ H.,
Alexander, D.\ R., Tamanai, A., \&  Schweitzer, A. 2001, \apj, 556, 357

\bibitem[Baraffe et al.(1997)]{bar97} Baraffe, I., Chabrier, G., Allard, F.,
\& Hauschildt, P.\ H. 1997, \aap, 327, 1054

\bibitem[Baraffe et al.(1998)]{bar98} Baraffe, I., Chabrier, G., Allard, F.,
\& Hauschildt, P.\ H. 1998, \aap, 337, 403
 
\bibitem[Baraffe et al.(2003)]{bar03}Baraffe, I., Chabrier, G., Barman, T., Allard, F.,
\& Hauschildt, P.\ H. 2003, \aap, 382, 563

\bibitem[Bildsten et al.(1997)]{bil97} Bildsten, L., Brown, E.\ F., Matzner, C.\ D., \& Ushomirsky, G. 1997, \apj, 482, 442

\bibitem[Borysow, J{\o}rgensen, \& Zheng(1997)]{bor97}Borysow, A., J{\o}rgensen,
U.\ G., \& Zheng, C. 1997, \aap, 324, 185

\bibitem[Boss(2006)]{bos06} Boss, A.\ P. 2006, \apj, 644, L79

\bibitem[Burgasser(2004)]{me1626}Burgasser, A.\ J. 2004, \apj, 614, L73

\bibitem[Burgasser, Cruz \& Kirkpatrick(2007)]{megmos}Burgasser, A.\ J.,
Cruz, K.\ L., \& Kirkpatrick, J.\ D. 2007, \apj, 657, 494

\bibitem[Burgasser et al.(2006)]{meclass2}Burgasser, A.\ J., Geballe, T.\ R.,
Leggett, S.\ K., Kirkpatrick, J.\ D., \& Golimowski, D.\ A. 2006, \apj, 637, 1067

\bibitem[Burgasser et al.(2003)]{me0532}Burgasser, A.\ J., Kirkpatrick,
J.\ D., Burrows, A., Liebert, J., Reid, I.\ N., Gizis, J.\ E., McGovern, M.\ R.,
Prato, L., \& McLean, I.\ S. 2003, \apj, 592, 1186

\bibitem[Burgasser, Kirkpatrick, \& L\'epine(2005)]{mecs13}Burgasser, A.\ J., Kirkpatrick, J.\ D., \& L\'epine, S. 2005 in The 13th Cambridge Workshop on Cool Stars, Stellar Systems, and the Sun (ESA-SP-560), ed.\ F.\ Favata, G.\ A.\ J.\ Hussain \& B.\ Battrick (Noordwijk: ESA), p.\ 237

\bibitem[Burgasser et al.(2007)]{meppv}Burgasser, A.\ J., Reid, I.\ N.,
Siegler, N., Close, L.\ M., Allen, P., Lowrance, P.\ J., \& Gizis, J.\ E.
2007, in Planets and Protostars V, eds.\ B.\ Reipurth, D.\ Jewitt and K.\ Keil (Univ.\ Arizona Press: Tucson), p.\ 427

\bibitem[Burrows et al.(2001)]{bur01}Burrows, A., Hubbard, W.\ B., Lunine,
J.\ I., \& Liebert, J. 2001, Rev.\ of Modern Physics, 73, 719

\bibitem[Burrows \& Sharp(1999)]{bur99}Burrows, A., \& Sharp, C.\ M. 1999, \apj,
512, 843

\bibitem[Burrows, Sudarsky \& Hubeny(2006)]{bur06}Burrows, A., Sudarsky, D., \& Hubeny, I.
2006, \apj, 640, 1063

\bibitem[Carney et al.(1996)]{car96}Carney, B.\ W., Laird, J.\ B., Latham, D.\ W., \& Aguilar, L.\ A. 1996, \aj, 112, 668

\bibitem[Carpenter(2001)]{car01}Carpenter, J.\ M. 2001, \aj, 121, 2851

\bibitem[Chiba \& Beers(2000)]{chi00}Chiba, M., \& Beers, T.\ C. 2000, \aj, 119, 2843

\bibitem[Cruz et al.(2003)]{cru03} Cruz, K.\ L., Reid, I.\ N., Liebert, J., Kirkpatrick, J.\ D.,
\& Lowrance, P.\ J. 2003, AJ, 126, 2421

\bibitem[Cruz et al.(2007)]{cru07} Cruz, K.\ L., et al. 2007, \aj, 133, 439

\bibitem[Cushing, Rayner \& Vacca(2005)]{cus05} Cushing, M.\ C., Rayner, J.\ T., \& Vacca, W.\ D. 2005, \apj, 623, 1115

\bibitem[Cushing et al.(2006)]{cus06}Cushing, M.\ C., et al. 2006, \apj, 648, 614

\bibitem[Dahn et al.(1995))]{dah95}Dahn, C.\ C., Liebert, J., Harris, H.\ C., \& Guetter, H.\ H. 1995 in The Bottom of the Main Sequence - and Beyond, ed.\ C.\ G.\ Tinney (Springer-Verlag: Heidelberg), p.239


\bibitem[Dahn et al.(2002)]{dah02}Dahn, C.\ C., et al. 2002, \aj,
124, 1170

\bibitem[D'Antona \& Mazzitelli(1985)]{dan85}D'Antona, F., \& Mazzitelli, I.
1985, \apj, 296, 502

\bibitem[Dauphole \& Colin(1995)]{dau95}Dauphole, B., \& Colin, J. 1995, \aap, 300, 117

\bibitem[Dehnen \& Binney(1998)]{deh98}Dehnen, W., \& Binney, J.\ J. 1998, \mnras,
298, 387

\bibitem[Digby et al.(2003)]{dig03}Digby, A.\ P., Hambly, N.\ C., Cooke, J.\ A., Reid, I.\ N.,
\& Cannon, R.\ D. 2003, \mnras, 344, 583

\bibitem[Epchtein et al.(1997)]{epc97}Epchtein, N., et al. 1997,
The Messenger, 87, 27

\bibitem[Fazio et al.(2004)]{faz04}Fazio, G.\ G., et al. 2004, \apjs, 154, 10

\bibitem[Fischer et al.(2003)]{fis03} Fischer, J., et al. 2003, Proc.\ SPIE, 4841, 564

\bibitem[Gizis(1997)]{giz97}Gizis, J.\ E. 1997, \aj, 113, 806


\bibitem[Golimowski et al.(2004)]{gol04} Golimowski, D.\ A., et al. 2004, \aj, 127, 3516

\bibitem[Guetter et al.(2003)]{gue03} Guetter, H.\ H., Vrba, F.\ J., Henden, A.\ A., \& Luginbuhl, C.\ B. 2003, \aj, 125, 3344

\bibitem[Hauschildt, Allard \& Baron(1999)]{hau99}Hauschildt, P.\ H., Allard, F.,
\& Baron, E. 1999, \apj, 512, 377

\bibitem[Helling et al.(2004)]{hel04}Helling, Ch., Klein, R., Woitke, R., Nowak, U., \& Sedlmayr, E. 2004, \aap, 423, 657

\bibitem[Kerr \& Lynden-Bell(1986)]{ker86}Kerr, F.\ J., \& Lynden-Bell, D.
1986, \mnras, 221, 1023

\bibitem[Kirkpatrick(2005)]{kir05}Kirkpatrick, J.\ D. 2005, \araa, 43, 195

\bibitem[Kirkpatrick et al.(2006)]{kir06}Kirkpatrick, J.\ D., Barman, T.\ S., Burgasser, A.\ J., McGovern, M.\ R., McLean, I.\ S., Tinney, C.\ G., \&
Lowrance, P.\ J. 2006, \apj, 639, 1120

\bibitem[Kirkpatrick et al.(1999)]{kir99}Kirkpatrick, J.\ D., et al. 1999,
\apj, 519, 802

\bibitem[Knapp et al.(2004)]{kna04}Knapp, G., et al. 2004, \apj, 127, 3553

\bibitem[Kuiper(1939)]{kui39} Kuiper, G.\ P. 1939, \apj, 89, 549


\bibitem[Laughlin, Bodenheimer \& Adam(1997)]{lau97} Laughlin, G., Bodenheimer, P., \& Adams, F.\ C. 1997, \apj, 482, 420

\bibitem[Leggett et al.(2000)]{leg00}Leggett, S.\ K., Allard, F., Dahn, C.,
Hauschildt, P.\ H., Kerr, T.\ H., \& Rayner, J. 2000, \apj, 535, 965

\bibitem[Leggett et al.(2007)]{leg07} Leggett, S.\ K., Saumon, D., Marley, M. S., Geballe, T.\ R., Golimowski, D.\ A., Stephens, D., \& Fan, X. 2007, \apj, 655, 1079

\bibitem[L\'epine et al.(2002)]{lep1826} L\'epine, S.,
Rich, R.\ M., Neill, J.\ D., Caulet, A., \& Shara, M.\ M. 2002, \apj, 581, L47

\bibitem[L\'epine, Rich, \& Shara(2003)]{lep1610} L\'epine, S.,
Rich, R.\ M., \& Shara, M.\ M. 2003, \apj, 591, L49

\bibitem[Liebert \& Burgasser(2007)]{lie07}Liebert, J., \& Burgasser, A.\ J. 2007, \apj, 655, 522

\bibitem[Linsky(1969)]{lin69}Linsky, J.\ L. 1969, \apj, 156, 989

\bibitem[Lodders(2002)]{lod02}Lodders, K. 2002, \apj, 577, 974

\bibitem[Luhman et al.(2007)]{luh07} Luhman, K.\ L., Joergens, V., Lada, C., Muzerolle, J., Pascucci, I., \& White, R. 2007, in Planets and Protostars V, eds.\ B.\ Reipurth, D.\ Jewitt and K.\ Keil (Univ.\ Arizona Press: Tucson), p.\ 443

\bibitem[Monet et al.(1992)]{mon92}Monet, D.\ G., Dahn, C.\ C., Vrba, F.\ J.,
Harris, H.\ C., Pier, J.\ R., Luginbuhl, C.\ B., \& Ables, H.\ D. 1992,
\aj, 103, 638

\bibitem[Mould (1976)]{mou76}Mould, J.\ R.  1976, \apj, 207, 535

\bibitem[Noll et al.(2000)]{nol00}Noll, K.\ S., Geballe, T.\ R., Leggett,
S.\ K., \& Marley, M.\ S. 2000, \apj, 541, L75

\bibitem[Norris(1994)]{nor94}Norris, J.\ E.  1994, \apj, 431, 645 

\bibitem[Patten et al.(2006)]{pat06}Patten, B.\ M., et al. 2006, \apj, 651, 502

\bibitem[Perryman et al.(1997)]{pry97}Perryman, M.\ A.\ C., et al. 1997,
\aap, 323, L49

\bibitem[Reid et al.(2000)]{rei00}Reid, I.\ N., Kirkpatrick, J.\ D.,
Gizis, J.\ E., Dahn, C.\ C., Monet, D.\ G., Williams, R.\ J.,
Liebert, J., \& Burgasser, A.\ J. 2000, \aj, 119, 369

\bibitem[Reid et al.(2002)]{rei02}Reid, I.\ N., Kirkpatrick, J.\ D.,
Liebert, J., Gizis, J.\ E., Dahn, C.\ C., \&
Monet, D.\ G. 2002, \aj, 124, 519

\bibitem[Reiners \& Basri(2006)]{rei06} Reiners, A., \& Basri, G. 2006, \aj, 131, 1806

\bibitem[Saumon et al.(1994)]{sau94}Saumon, D., Bergeron, P., Lunine, J.\ I.,
Hubbard, W.\ B., \& Burrows, A. 1994, \apj, 424, 333

\bibitem[Schmidt et al.(2007)]{sch07}Schmidt, S.\ J., Cruz, K.\ L., Bongiorno,
B.\ J., Liebert, J., \& Reid, I.\ N. 2007, \aj, in press

\bibitem[Scholz et al.(2004)]{sch1444}Scholz, R.-D., Lodieu, N., Ibata, R., Bienaym\'{e}, O.,
Irwin, M., McCaughrean, M.\ J., \& Schwope, A. 2004, \mnras, 347, 685

\bibitem[Sivarani, Kembhavi \& Gupchup(2007)]{siv07} Sivarani, T., Kembhavi, A.\ K., \& 
Gupchup, J. 2007, \apj, in preparation

\bibitem[Skrutskie et al.(2006)]{skr06}Skrutskie, M.\ F., et al. 2006, \aj, 131, 1163

\bibitem[Smith et al.(2003)]{smi03}	Smith, V.\ V., Tsuji, T., Hinkle, K.\ H., Cunha, K., Blum, R.\ D., Valenti, J.\ A., Ridgway, S.\ T., Joyce, R.\ R., \& Bernath, P. 2003, \apj, 599, L107

\bibitem[Sommer-Larsen \& Zhen(1990)]{som90} Sommer-Larsen, J., \& Zhen, C. 1990, \mnras, 242, 10

\bibitem[Stassun et al.(2006)]{sta06}Stassun, K., Mathieu, R.\ D., Vaz, L.\ P.\ R., 
Valenti, J.\ A., \& Gomez, Y. 2006, Nature, 440, 311

\bibitem[Stauffer, Schultz, \& Kirkpatrick(1998)]{sta98}Stauffer, J.\ R.,
Schultz, G., \& Kirkpatrick, J.\ D. 1998, \apj, 499, 199

\bibitem[Tarter et al.(2007)]{tar07}Tarter, J.\ C., et al. 2007, AsBio, 7, 30

\bibitem[Tinney, Burgasser, \& Kirkpatrick(2003)]{tin03}Tinney, C.\ G., Burgasser, A.\ J.,
\& Kirkpatrick, J.\ D. 2003, \aj, 126, 975

\bibitem[Tsuji, Ohnaka, \& Aoki(1996)]{tsu96}Tsuji, T., Ohnaka, K., \&
Aoki, W. 1996, \aap, 305, L1

\bibitem[Tsuji, Ohnaka, \& Aoki(1999)]{tsu99} ---. 1999, \apj, 520, L119

\bibitem[Vrba et al.(2004)]{vrb04}Vrba, F.\ J., et al. 2004, \aj, 127, 2948

\bibitem[York et al.(2000)]{yor00} York, D.\ G., et al. 2000, AJ,
120, 1579

\end{thebibliography}
\end{document}